\documentclass[doublecol]{epl2} 

\usepackage{amssymb}
\usepackage{amsmath}
\usepackage{graphicx}
\usepackage{amsbsy}
\usepackage{color}

\newcommand{\bq}{\begin{eqnarray}}
\newcommand{\eq}{\end{eqnarray}}
\newcommand{\bqn}{\begin{eqnarray*}}
\newcommand{\eqn}{\end{eqnarray*}}

\newcommand\beq{\begin{equation}}
\newcommand\eeq{\end{equation}}
\newcommand\beqa{\begin{eqnarray}}
\newcommand\eeqa{\end{eqnarray}}

\title{Phase diagram of the penetrable square well-model}

\author{Riccardo Fantoni\inst{1} \and Alexandr Malijevsk\'y\inst{2}
  \and Andr\'es Santos\inst{3} \and Achille Giacometti\inst{4}}

\institute{
\inst{1} National Institute for Theoretical Physics (NITheP) and
  Institute of Theoretical Physics, Stellenbosch 7600, South Africa \\
\inst{2} E. H\'ala Laboratory of Thermodynamics, Institute of
  Chemical Process Fundamentals of the ASCR, and Department of
  Physical Chemistry, Institute of Chemical Technology, Prague, 166 28
  Praha 6, Czech Republic \\
\inst{3} Departamento de F\'isica, Universidad de Extremadura,
  E-06071 Badajoz, Spain \\
\inst{4} Dipartimento di Chimica Fisica, Universit\`a Ca' Foscari Venezia,
  S. Marta DD2137, I-30123 Venezia, Italy
}

\abstract{
We study a system formed by soft colloidal spheres attracting each other
via a square-well potential, using extensive Monte Carlo simulations
of various nature. 
The softness is implemented through a reduction of the infinite
part of the repulsive potential 
to a finite one. For sufficiently low values of the penetrability parameter
we find the system to be Ruelle stable with square-well like behavior. For
high values of the penetrability the system is thermodynamically
unstable and collapses into an isolated blob formed by a few
clusters each containing many overlapping particles. For intermediate values of
the penetrability the system has a rich phase diagram with a partial
lack of thermodynamic consistency. 
}


\begin{document}
\maketitle
Pair effective interactions in soft condensed matter physics can be of
various nature and one can {often} find real systems
whose interaction is bounded at small separations as, {for
instance}, in the case of star and chain polymers \cite{Likos2001}.
In this case, paradigmatic models, such as square-well (SW)
fluids, that have been 
rather successful in predicting thermo-physical properties of simple
liquids, are no longer useful. Instead, different minimal models
accounting for the boundness of the 
potential have to be considered, the Gaussian core model  
\cite{Stillinger1976} and the penetrable-sphere (PS) model
\cite{Likos1998,Marquest1989,Malijevsky2006} 
being well studied examples. More recently, the penetrable square-well
(PSW) fluid has been added 
to this category \cite{Santos2008,Fantoni2009,Fantoni2010,Fantoni2010b} with
the aim of including the existence of attractive effective potentials.
The PSW model is obtained from the SW potential by
reducing to a finite value the infinite repulsion at short range,
\bq
\phi_{\text{PSW}}(r)=\left\{
\begin{array}{ll}
\epsilon_r ~,& r\le\sigma~,\\
-\epsilon_a ~,& \sigma<r\le\sigma+\Delta~,\\
0           ~,& r>\sigma+\Delta~,
\end{array}\right.
\eq
where $\epsilon_r$ and $\epsilon_a$ are two positive {energies}
accounting for the repulsive and attractive parts of the potential,
respectively, $\Delta$ is the width of the attractive square well, and
$\sigma$ is the width of the repulsive barrier. For $\epsilon_r\to\infty$
one recovers the SW model, while for $\Delta=0$ or
$\epsilon_a=0$ one recovers the PS model.

{For finite $\epsilon_a$, $\epsilon_a/\epsilon_r$ is a
measure of the penetrability of the barrier and we shall refer to
$\epsilon_a/\epsilon_r$ as the penetrability
ratio. PSW pair potentials can be obtained as effective potentials
for instance in polymer mixtures\cite{Bolhuis2001,McCarty2010}. While
in the majority of the cases the well depth $\epsilon_a$ is much
smaller than the repulsive barrier $\epsilon_r$ (low penetrability
limit) this mesoscopic objects are highly sensitive to external
conditions (e.g. quality of the solvent) and may thus in principle
exhibit higher values of the penetrability ratio
$\epsilon_a/\epsilon_r$.}  

It is well known that three-dimensional SW fluids
exhibit a fluid-fluid phase transition for any width of the
attractive square well 
\cite{Vega1992,deMiguel1997,delRio2002,Liu2005,Giacometti2009}, the
liquid phase becoming metastable against the formation of the solid
for a sufficiently narrow well \cite{Liu2005}.
It is also well established that in the PS fluid (that lacking an
attractive component in the pair potential cannot have a fluid-fluid
transition) an increase of the density leads to the formation of
clusters of overlapping particles arranged in an ordered crystalline
phase \cite{Klein1994,Likos1998,Schmidt1999,Mladek2008}. 

While the novel features appearing even in the one-dimensional case
have been studied in some details
\cite{Santos2008,Fantoni2009,Fantoni2010,Fantoni2010b}, no analysis
regarding the influence of penetrability and attractiveness on the
phase behavior of PSW fluids have been reported, so far, in three
dimensions. The present paper aims to fill this vacancy.

{A system is defined to be Ruelle stable
when the total potential energy $U_N$ for a
system of $N$ particles satisfies the condition $U_N\ge -NB$ where $B$
is a finite positive constant \cite{Ruelle,Fisher1966}.
In Ref.\ \cite{Fantoni2010} we proved that in the one-dimensional case
the PSW model is Ruelle stable if
$\epsilon_a/\epsilon_r<1/2(\ell+1)$, 
where $\ell$ is the integer part of $\Delta/\sigma$. This result can
be extended to any dimensionality $d$ by the following arguments. The
configuration which minimizes the energy of $N$ particles interacting
via the PSW potential is realized when $M$ closed packed clusters,
each consisting of $s=N/M$ particles collapsed into one point, are
distributed such that the distance between centers of two neighbor
clusters is $\sigma$. In such a configuration, all the
particles of the same cluster interact repulsively, so the repulsive
contribution to the total potential energy is $\epsilon_r
Ms(s-1)/2$. In addition, the particles of a given cluster interact 
attractively with all the particles of those $f_\Delta$ 
clusters within a distance smaller that $\sigma+\Delta$. In the
two-dimensional case, $f_\Delta=6$ and $12$ if
$\Delta/\sigma<\sqrt{3}-1$ and $\sqrt{3}-1<\Delta/\sigma<1$,
respectively. For $d=3$, the case we are interested in, one has
$f_\Delta=12$, $18$, and $42$ if $\Delta/\sigma<\sqrt{2}-1$,
$\sqrt{2}-1<\Delta/\sigma<\sqrt{3}-1$, and
$\sqrt{3}-1<\Delta/\sigma<1$, respectively. The attractive
contribution to the total potential energy is thus
$-\epsilon_a(M/2)\left[{f_\Delta}-b_\Delta(M)\right]s^2$, where
$b_\Delta(M)$  accounts for a reduction of the actual number of
clusters interacting attractively, due to boundary effects. This
quantity has the properties $b_\Delta(M)<{f_\Delta}$, $b_\Delta(1)=
{f_\Delta}$, and $\lim_{M\to\infty}b_\Delta(M)=0$. For instance, in
the two-dimensional case with $\Delta/\sigma<\sqrt{3}-1$ one has
$b_\Delta(M)=2(4\sqrt{M}-1)/M$.  Therefore, the total potential energy
is 
\bq
\frac{U_N(M)}{N\epsilon_r}=-\frac{1}{2}+\frac{N}{2M}F(M),
\eq
where $F(M)\equiv (\epsilon_a/\epsilon_r)b_\Delta(M)+\left(1-f_\Delta
\epsilon_a/\epsilon_r\right)$.  
If $\epsilon_a/\epsilon_r<1/f_\Delta$, $F(M)$ is positive definite, so
$U_N(M)/N$ has a lower bound and the system is stable in the
thermodynamic limit. On the other hand, If
$\epsilon_a/\epsilon_r>1/f_\Delta$, one has $F(1)=1$ but
$\lim_{M\to\infty}F(M)=-(f_\Delta \epsilon_a/\epsilon_r-1)<0$. In that
case, there must exist a certain \emph{finite} value $M=M_0$ such that
$F(M)<0$ for $M>M_0$. As a consequence, in those configurations with
$M>M_0$, $U_N(M)/N$ has no lower bound in the limit $N\to\infty$ and
thus  the system may be unstable.} 

\begin{figure}[ht!]
\begin{center}
\includegraphics[width=8cm]{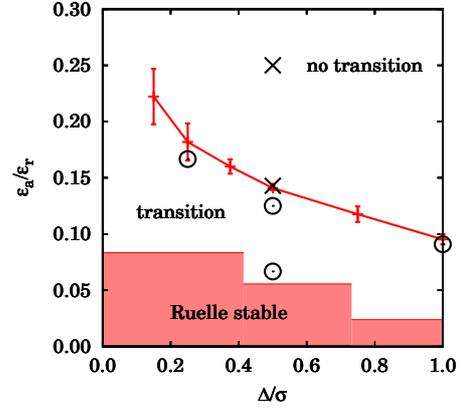}
\end{center}
\caption{Plot of the penetrability ratio $\epsilon_a/\epsilon_r$ as a
function 
of $\Delta/\sigma$. The displayed line separates the parameter region
where the PSW model admits a fluid-fluid phase transition from that
where it does not. {The highlighted region ($\epsilon_a/\epsilon_r\leq
1/12$ for $\Delta/\sigma<\sqrt{2}-1$, $\epsilon_a/\epsilon_r\leq
1/18$ for $\sqrt{2}-1<\Delta/\sigma<\sqrt{3}-1$, and $\epsilon_a/\epsilon_r\leq
1/42$ for $\sqrt{3}-1<\Delta/\sigma<1$) shows where  
the model is expected to be thermodynamically stable in the
sense of Ruelle for any thermodynamic state}. The SW
model falls on the horizontal axis ($\epsilon_a/\epsilon_r\to 0$) 
and its fluid-fluid transition is expected to be metastable 
against the freezing transition for $\Delta/\sigma\lesssim 0.25$
\cite{Liu2005}. The circles are the points chosen for the calculation
of the coexistence lines (see Figs. \ref{fig:gl} and
\ref{fig:isotherm}). The crosses are the points chosen for the 
determination of the boundary between extensive and non-extensive
phases (see Fig. \ref{fig:inst}).}  
\label{fig:pd}
\end{figure}

We have performed an extensive analysis of the vapor-liquid phase
transition of the system using Gibbs Ensemble Monte Carlo (GEMC) simulations
\cite{Frenkel-Smit,Panagiotopoulos87,Panagiotopoulos88,Smit89a,Smit89b},
starting from the corresponding SW fluid condition and gradually
increasing the penetrability {ratio $\epsilon_a/\epsilon_r$}
until the transition disappears.  
A total number of $N=512$ particles with $2N$ particle random 
displacement, $N/10$ volume changes, and $N$ particle swap moves
between the gas and the liquid box,
{on average per cycle}, were considered.
{We find that for any given width $\Delta/\sigma<1$ of the
well, there is a limit value of the penetrability ratio
$\epsilon_a/\epsilon_r$ above which no fluid-fluid phase transition is
observed. This is depicted in Fig. \ref{fig:pd} where it can be
observed that (for $\Delta/\sigma<1$) this line lies outside the
Ruelle stable region $\epsilon_a/\epsilon_r<1/f_\Delta$.}

\begin{figure}[ht!]
\begin{center}
\includegraphics[width=8cm]{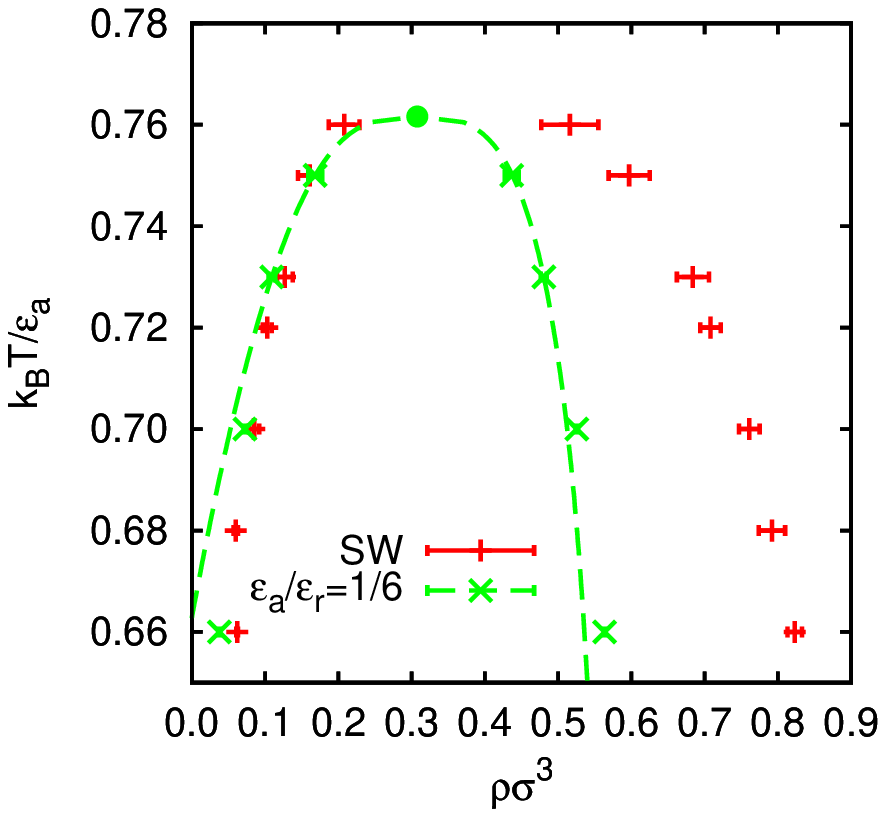}\\
\includegraphics[width=8cm]{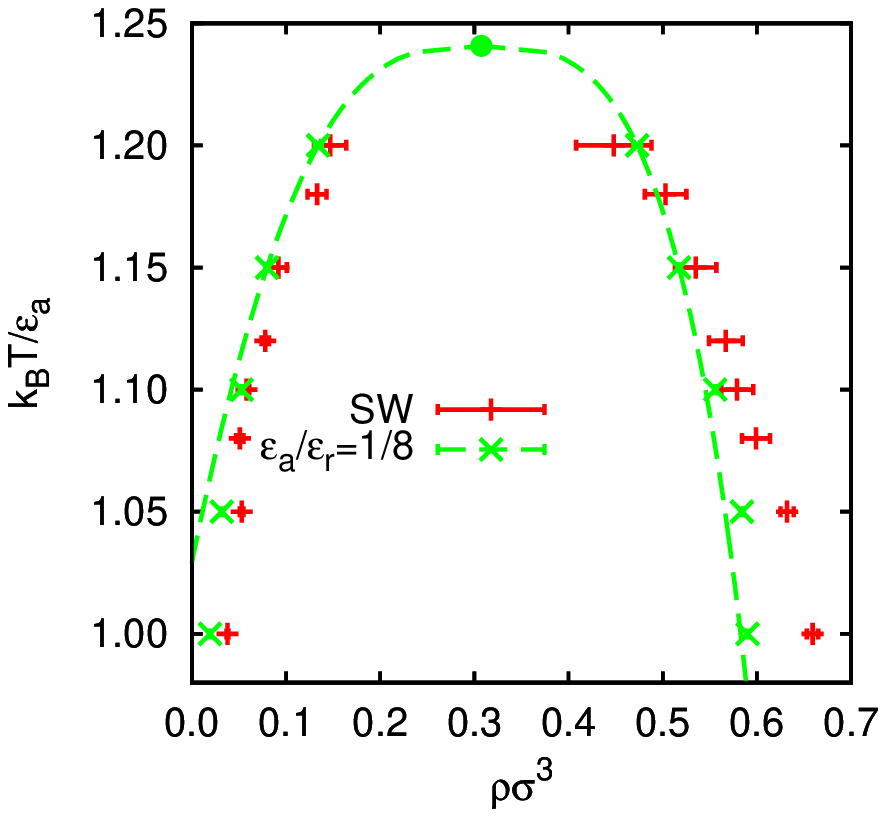}\\
\includegraphics[width=8cm]{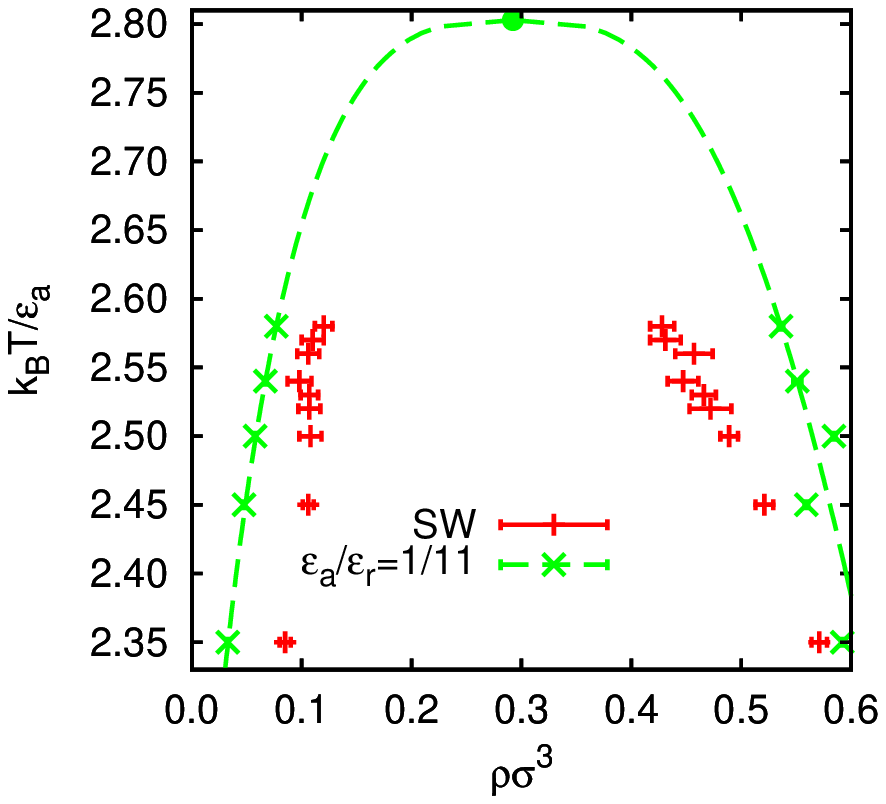}
\end{center}
\caption{Fluid-fluid coexistence line. The solid circle represents the
critical point $(\rho_c,T_c)$. In the top panel, for 
$\Delta/\sigma=0.25$ and $\epsilon_a/\epsilon_r=1/6$, one has
$\rho_c\sigma^3=0.307$ and $k_BT_c/\epsilon_a=0.762$; in the middle
panel, below the limit penetrability,
for $\Delta/\sigma=0.5$ and $\epsilon_a/\epsilon_r=1/8$, one has
$\rho_c\sigma^3=0.307$ and $k_BT_c/\epsilon_a=1.241$; and in the
bottom panel for $\Delta/\sigma=1.0$ and
$\epsilon_a/\epsilon_r=1/11$, one has $\rho_c\sigma^3=0.292$ and
$k_BT_c/\epsilon_a=2.803$. The lines are the result of the fit with 
the law of rectilinear diameters. The SW results are the ones of
Vega et al.  \cite{Vega1992}}
\label{fig:gl}
\end{figure}

{It is instructive to analyze the detailed form of the
coexistence curves below (but close to) the limit line of
Fig. \ref{fig:pd}. This is presented in  
Fig. \ref{fig:gl}}. We have explicitly checked that our code
reproduces completely the 
results of Vega et al. \cite{Vega1992} for the SW
model. {Following standard procedures \cite{Vega1992}} we
fitted the GEMC points, near the critical point using 
the law of rectilinear diameters $(\rho_l+\rho_g)/2=\rho_c+A(T_c-T)$
where $\rho_l$ $(\rho_g)$ is the density of the liquid (gas) phase,
$\rho_c$ is the critical density, and $T_c$ is the critical
temperature. Furthermore, the temperature dependence of the density
difference of the coexisting phases is fitted to the scaling
law $\rho_l-\rho_g=B(T_c-T)^\beta$
where $\beta=0.32$ is the critical exponent for the three-dimensional
Ising model. The amplitudes $A$ and $B$ where determined from the fit.
{In the state points above the limit line of
Fig. \ref{fig:pd} we have considered temperatures below the critical
temperature of the corresponding SW system. The disappearance of the
fluid-fluid transition is signaled by the evolution towards an empty
gas box and a clustered phase in the liquid box.}

\begin{figure}[ht!]
\begin{center}
\includegraphics[width=8cm]{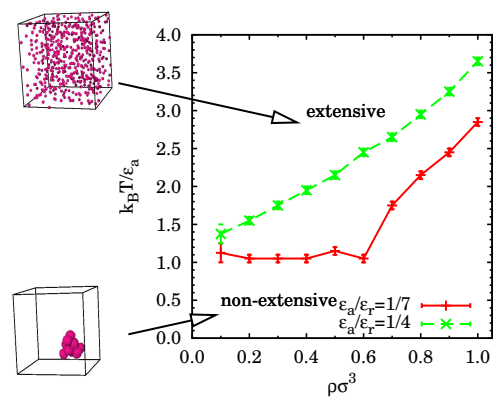}
\end{center}
\caption{Regions of the phase diagram where the PSW fluid,
with $\Delta/\sigma=0.5$ and two different values of 
$\epsilon_a/\epsilon_r$, is found to be extensive or non-extensive
(here we used $N=512$ particles). The left shows representative snapshots
in the two regions. {The
instability line corresponding to the higher penetrability case
(dashed line) lies above the one corresponding to lower penetrability
(continuous line).}} 
\label{fig:inst}
\end{figure}

As discussed, the PSW fluid is thermodynamically Ruelle stable when  
{$\epsilon_a/\epsilon_r<1/f_\Delta$} for all values of the
thermodynamic 
parameters. {For $\epsilon_a/\epsilon_r>1/f_\Delta$ the
system is either extensive or non-extensive
depending on temperature and density. For a given density, one could
then expect that there exists a certain temperature $T_\text{inst}(\rho)$,
such that the system is metastable if $T>T_\text{inst}$ and unstable if
$T<T_\text{inst}$}.  
{We determined the metastable/unstable crossover by
performing NVT Monte Carlo simulations with $N=512$
particles initially uniformly distributed within the simulation
box. Figure \ref{fig:inst} reports the results in the reduced
temperature-density plane for $\Delta/\sigma=0.5$ and for two selected
penetrability ratios $\epsilon_a/\epsilon_r=1/7$ and $1/4$, the first
one lying exactly just above the limit line of Fig. \ref{fig:pd}
while the second deep in the non-transition region.} We worked with
constant size moves (instead of fixing the acceptance ratios) during
the simulation run, choosing the move size of $0.15$ in units of the 
the simulation box side. 
{A crucial point in the above numerical analysis is the
identification of the onset of the instability. Clearly the physical
origin of this instability stems from the fact that the attractive
contribution increases unbounded compared to the repulsive one and
particles tend to lump up into clusters of multiply overlapping
particles (``blob''). Hence the energy can no longer scale linearly
with the total number of particles $N$ and the thermodynamic limit is
not well defined (non-extensivity). We define a cluster in the following 
way. Two particles belong to the same cluster if there
is a path connecting them, where we are allowed to move on a path
going from one particle to another if the centers of the two particles
are at a distance less than $\sigma$.}

The state points belonging to the unstable region are characterized by
a sudden drop of the internal energy and of the acceptance ratios at some
points in the system evolution during the MC
simulation. Representative snapshots show that a blob 
structure has nucleated around a certain 
point and occupies {only a part of} the simulation box with a
few clusters. {The number of clusters decreases as one 
moves away from the boundary line found in Fig. \ref{fig:inst} towards
lower temperatures}. Upon increasing $\epsilon_a/\epsilon_r$ 
{the number of clusters decreases and the
number of particles per cluster increases}. We assume 
a state to be {metastable} if the energy does 
not undergo the transition after $10^7 N$ single particle moves.  
The fluid-fluid transition above the limit
penetrability line of Fig. \ref{fig:pd} is not possible because the
non-extensive phase shows up before the critical point is reached.

The boundary line of Fig. \ref{fig:inst} is robust with respect to the
size of the system, provided that a
sufficiently large size ($N \ge 512$) is chosen. When the number of
particles in the 
simulation goes below the number of clusters which would form in the
non-extensive phase the system seems to remain extensive. For instance,
with $N=1024$ particles we obtained under the $\epsilon_a/\epsilon_r=1/7$,
$\Delta/\sigma=0.5$ conditions a threshold temperature
$k_BT/\epsilon_a\approx 1.15$ for $\rho\sigma^3=0.4$ and  
$2.25$ for $\rho\sigma^3=0.8$, which are close to the values
obtained with $N=512$ particles. 

There is an {apparent} hysteresis in forming and melting the
non-extensive phase. For example when $\epsilon_a/\epsilon_r=1/7$,
$\Delta/\sigma=0.5$, and $\rho\sigma^3=1.0$ the non-extensive phase starts
forming when cooling down to $k_BT/\epsilon_a=2.75$. Upon increasing
the temperature 
again, we observed a melting transition at significantly higher
temperatures ($k_BT/\epsilon_a\gtrsim 4$). We also 
found the hysteresis to be size dependent; in the same state for
$\rho\sigma^3=0.6$ the melting temperatures are
$k_BT/\epsilon_a\approx 2.5$ for $N=256$, $k_BT/\epsilon_a\approx 4.5$
for $N=512$, and 
$k_BT/\epsilon_a\approx 6.5$ for $N=1024$. This suggests that the extensive
phase in Fig. \ref{fig:inst} is actually metastable with respect to
the non-extensive phase in the thermodynamic limit. However the
metastable phase can be stabilized by taking the size of the system
finite. {In addition we cannot exclude, a priori, the
possibility of a true extensive stable phase as it is not prevented by
the Ruelle criterion. We note that the size dependence of the hysteresis 
in the melting could be attributed to the fact that the blob occupies
only part of the simulation box and therefore a surface term has a
rather high impact on the melting temperature.}

{A convenient way to characterize the structure of the fluid
is to consider the radial distribution function $g(r)$. This is
depicted in Fig. \ref{fig:gs} for the cases
$\Delta/\sigma=0.5$, $k_BT/\epsilon_a=1.20$, and  
$\rho\sigma^3=0.7$ at $\epsilon_a/\epsilon_r=1/8$ and
$\epsilon_a/\epsilon_r=1/7$. The latter case is in the 
non-extensive region, according to Fig. \ref{fig:inst}. 
\begin{figure}[h!]
\begin{center}
\includegraphics[width=8cm]{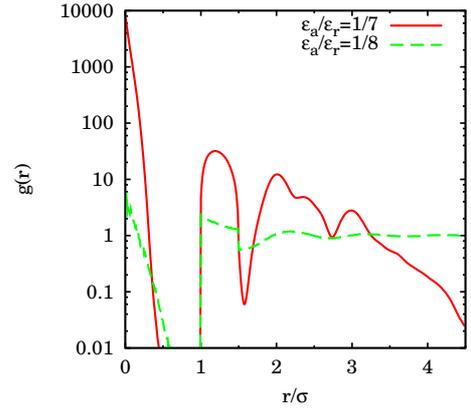}
\end{center}
\caption{Radial distribution function for the PSW model at
$\Delta/\sigma=0.5$, $k_BT/\epsilon_a=1.20$, and $\rho\sigma^3=0.7$
for two different values of the penetrability ratio
$\epsilon_a/\epsilon_r$.} 
\label{fig:gs}
\end{figure}
We can clearly see that there is a dramatic change in the structural
properties of the PSW liquid. In the non-extensive case,
$\epsilon_a/\epsilon_r=1/7$, the radial 
distribution function grows a huge peak at $r=0$ and decays to zero
after the first few peaks, which suggests clustering and confinement
of the system.} 

In order to study the solid phase of the PSW model below the limit
penetrability we employed isothermal-isobaric (NPT) MC
simulations. A typical run would consist of $10^8$ steps (particle
moves or volume moves) with an
equilibration time of $10^7$ steps. We used $108$ particles and
adjusted the particle moves to have acceptance ratios of $\approx 0.5$
and volume changes to have acceptance ratios of $\approx 0.1$.
Here we only consider the case of PSW with $\Delta/\sigma=0.5$ and
$\epsilon_a/\epsilon_r=1/8$ and $1/15$. 

\begin{figure}[ht!]
\begin{center}
\includegraphics[width=8cm]{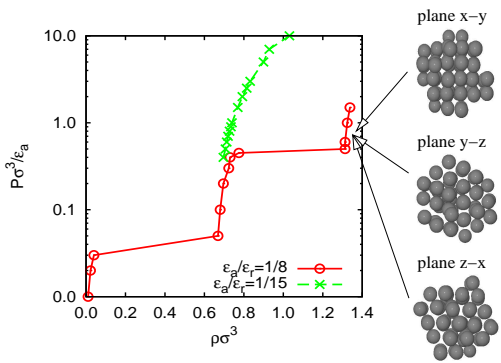}
\end{center}
\caption{Isotherm $k_BT/\epsilon_a=1$ for the PSW system with
$\Delta/\sigma=0.5$ and $\epsilon_a/\epsilon_r=1/8$ and
$\epsilon_a/\epsilon_r=1/15$, as obtained from NPT MC simulation with
$N=108$ particles. The pressure axis is in logarithmic scale. The
right shows snapshots of the centers of mass of the clusters in the
solid.}  
\label{fig:isotherm}
\end{figure}

For the SW system with a width $\Delta/\sigma=0.5$ the
critical point is known to be at $k_BT_c/\epsilon_a=1.23$ and
$\rho_c\sigma^3=0.309$, its triple point being at
$k_BT_t/\epsilon_a=0.508$, $P_t\sigma^3/\epsilon_a=0.00003$,
$\rho_l\sigma^3=0.835$, and $\rho_s\sigma^3=1.28$ \cite{Liu2005}. No
solid stable phase was found in Ref. \cite{Liu2005} for temperatures
above the triple point, meaning that the 
melting curve in the pressure-temperature phase diagram is almost vertical.
On the other hand, the phase diagram of the PSW fluid with the same
well width and a 
value of $\epsilon_a/\epsilon_r=1/8$, just below the limit line of
Fig. \ref{fig:pd}, shows that the melting curve has a smooth positive
slope in the pressure-temperature phase diagram.
In order to establish this, we used NPT simulations to
follow the $k_BT/\epsilon_a=1$ isotherm. From 
Fig. \ref{fig:isotherm} we can clearly see the 
jumps in density corresponding to the gas-liquid and to the
liquid-solid coexistence regions.  
The presence of a solid phase can be checked
by computing the $Q_6$ order parameter \cite{Wolde1996}{,
calculated for 
the center of mass of individual clusters,} that in
the present case turns out to be $Q_6\approx 0.35$. The crystal
structure is triclinic with a unit cell with $a=b=c=\sigma$ and
$\alpha=\beta=\pi/3$ and $\gamma=\cos^{-1}(1/4)$. There are
possibly other solid-solid coexistence regions at 
higher pressures. Moreover the relaxation time of the MC run in the
solid region is an order of magnitude higher than the one in the
liquid region. 

We also run at the same temperature a set of simulations for the PSW
fluid with $\epsilon_a/\epsilon_r=1/15$. The results (see 
Fig. \ref{fig:isotherm}) showed no indication of a
stable solid, in agreement with the fact that at this very low value of
the penetrability ratio the system is SW-like.

A peculiarity of the PSW in the region below the limit penetrability
of Fig. \ref{fig:pd}, but 
not in the Ruelle stability region, is a violation of the
Clausius-Clapeyron equation \cite{Kofke1993b} along the liquid-solid
coexistence curve, 
what represents a partial lack of thermodynamic consistency 

In the intermediate penetrability case (i.e., above Ruelle's threshold
but below the limit penetrability), the observed crystal
structure is made of clusters of overlapping particles {(rarely more
than two)} located at the sites of a regular crystal 
lattice. It is precisely this additional degree of penetrability,
not present in the SW system, that allows for the coexistence of the
liquid and the solid at not excessively large pressures. {In
this respect qualitative arguments along the lines suggested in
Ref. \cite{Mladek2008} could be useful}.

In conclusion, we have studied the phase diagram of the three-dimensional
PSW system. This model is Ruelle stable for
{$\epsilon_a/\epsilon_r<1/f_\Delta$}. For 
{$\epsilon_a/\epsilon_r>1/f_\Delta$} is either metastable or unstable
(non-extensive),  
depending on the values of temperature and density, as shown in
Fig. \ref{fig:inst}. The instability is indicated by the collapse of
the system in a confined blob made
up of a few clusters of several overlapping particles. Moreover, the
gas-liquid phase transition disappears, as shown in Fig. \ref{fig:pd}.

For the metastable fluid near the limit penetrability line of
Fig. \ref{fig:pd} we determined the 
phase diagram comparing it with the corresponding SW case. We
determined how the gas-liquid coexistence curves are
modified by the presence of penetrability (see Fig. \ref{fig:gl}) and
discussed the main features of the phase diagram, including the solid
phase, for $\Delta/\sigma=0.5$.

For the liquid-solid coexistence curves we generally found that the
solid density increases with respect to the corresponding SW
case, as expected, due to the formation of clusters of overlapping
particles in the crystal sites. For
$\Delta/\sigma=0.5$ the PSW model with a sufficient penetrability to
have a metastable system, but not a Ruelle stable one, has a melting curve
with a positive slope in the
pressure-temperature phase diagram with a violation of the
Clausius-Clapeyron thermodynamic equation, thus confirming the
metastable character of the phases. For sufficient low penetrability 
the system is in the Ruelle stable region, and behaves as
the corresponding SW model.

In summary, by experimentally tuning the repulsive barrier relative to
the well depth one could observe (a) stable phases resembling those of
a normal fluid, (b) metastable phases with fluid-fluid and fluid-solid
coexistence, or (c) the collapse of the system to a small region. 

\acknowledgments
We thank Tatyana Zykova-Timan and Bianca M. Mladek for enlighting
discussions and useful suggestions.
The support of PRIN-COFIN 2007B58EAB, FEDER
FIS2010-16587 (Ministerio de Ciencia e Innovaci\'on), GAAS
IAA400720710 is acknowledged.  
\bibliographystyle{eplbib}
\bibliography{3dpsw}

\end{document}